\documentclass[12pt]{article}
\usepackage{amsfonts}
\usepackage{amssymb}

\def \id {1\!\!{\rm I}}

\def\cint{\ifinner{\int\hspace*{-0.35cm}-}\else{\int\hspace*{-0.4cm}-}\fi}

\begin{document}
\baselineskip=.6cm

\begin{center}
{\Large{\bf On The Construction of Zero Energy States\\
in Supersymmetric Matrix Models III}}
\end{center}

\vspace{3cm}

\begin{center}
{\large Jens Hoppe$^*$}\\
{\large Theoretische Physik, \ ETH-H\"onggerberg}\\
{\large CH-8093 \ Z\"urich, Switzerland}
\end{center}

\vspace{6cm}

\noindent
{\bf Abstract}\\
For a supersymmetric Hamiltonian appearing in the matrix model related
to 11 dimensional supermembranes, zero energy states are
constructed. A useful symmetry, and an energy-equipartition property
is pointed out.

\vspace{8cm}

{\small $^*$ Heisenberg fellow

\vfill\eject

\noindent The supercharges of the previously studied SU(N) matrix
model for supermembranes in 11 space-time dimensions are sums of two
terms (each being, separately, nilpotent and defining it's own
supersymmetric Hamiltonian):
\begin{eqnarray}
Q_\beta &=&
M_a^{(\beta)}\,\lambda_a\;+\;D_a^{(\beta)}\,\partial_{\lambda_a}\nonumber \\
Q_\beta^\dagger &=& D_a^{(\beta)\dagger}\,\lambda_a\;+\; M_a^{(\beta)\dagger}\,
\partial_{\lambda_a}\\
&&\phantom{mmmmmn} \beta\,=\,1,\ldots, 8 \nonumber
\end{eqnarray}
where $\{ \lambda_a, \lambda_b\} = 0 = \{ \partial_{\lambda_a},
\partial_{\lambda_b}\}$ , $\{ \lambda_a,
\partial_{\lambda_b}\}=\delta_{ab}$, and the $M_a$ and $D_a$ ,
$a=(\alpha, A)=(1,1) 
\ldots (8, N^2-1)$, are first order differential operators (with
respect to bosonic variables $(x_{jA}, z_A,
\overline{z}_A)_{j=1\ldots 7 \phantom{mm}\atop A=1\ldots N^2-1}$) ,
given explicitly in terms of purely imaginary, antisymmetric,
$8\times8,$ \ SO(7) Gamma matrices by
\begin{eqnarray}
&&M_{\alpha A}^{(\beta)}\;=\;\delta_{\alpha\beta}\; i\,q_A
\;+\;\Gamma_{\alpha\beta}^j \ \frac{\partial}{\partial
  x_{jA}}\;-\;\frac 1 2 \ f_{ABC}\; x_{jB}\; x_{kC}\;
\Gamma_{\alpha\beta}^{jk}\nonumber\\ 
&&D_{\alpha A}^{(\beta)}\;=\;\delta_{\alpha\beta}\,2
\partial_A\,-\,i\, f_{ABC}\; x_{jB} \;\overline{z}_C\;\Gamma_{\alpha
  \beta}^j \ ,
\end{eqnarray}
with $q_A\,:=\,\frac i 2 \ f_{ABC}\,z_B\overline{z}_C$ \ (and
$f_{ABC}$ being real, totally antisymmetric structure constants of
SU(N)).

\medskip

\noindent In [1], solutions of $Q_\beta \chi = 0$ , $Q_\beta^\dagger \Psi =
0$, were constructed that are of the form
\begin{eqnarray}
\Psi &=& (\id - A)^{-1} \ \Psi^{(h)} \nonumber\\
\chi &=& (\id - B)^{-1} \ \chi^{[h]}
\end{eqnarray}
with 
\begin{eqnarray}
&& A\;=\; \left(I^\dagger \cdot \lambda\right) \left( D^\dagger \cdot
  \lambda\right) \ , \ B\;=\; \left( I \cdot \partial_\lambda\right) \left(
  D\cdot \partial_\lambda\right)\nonumber\\
&&\phantom{mmmmmmm} I_{\alpha A}^{(\beta)} \; :=\; i\,
\delta_{\alpha\beta} \ \frac{q_A}{q^2} \ , \\
{\rm and}\phantom{mmmmmmmmm} &&\left( M^\dagger\cdot \partial_\lambda\right)\,
\Psi^{(h)} \;=\; 0 \ , \ \left( M\cdot \lambda\right)\;\chi^{[h]}\;=\;0
\ . \phantom{mmmmmmmmmmmmmm}\nonumber
\end{eqnarray}
Here, I will only consider 
\begin{equation}
Q_D^{(\beta)} \; :=\; D_a^{(\beta)} \ \partial_{\lambda_a} \ ,
\end{equation}
satisfying $\{ Q_D^{(\beta)}\; , \; Q_D^{(\beta')}\}=0$ , and giving
rise to the non-negative supersymmetric quantum-mechanical Hamiltonian
$Q_\beta\,Q_\beta^\dagger\,+\,Q_\beta^\dagger\,Q_\beta$ , which, in a
particular representation for the $\Gamma$ matrices, reads 
\begin{eqnarray}
H_D\;=\;&-& 4\, \partial_A\;
\overline{\partial}_A + f_{ABC}\;x_{jB}\;\overline{z}_C\;
f_{AB'C'}\;x_{jB'}\;z_{C'} \nonumber\\ 
&+& 2\, f_{AA'E}\;x_{jE} \left( \delta_{\alpha 8}\;\delta_{\alpha j'}-
  \delta_{\alpha j}\;\delta_{\alpha'8}\right)\;\lambda_{\alpha A}\;
\partial_{\lambda_{\alpha' A'}}
\end{eqnarray}
(for notational simplicity, $\beta$ is chosen to be 8).

Despite the fact that (6) does not have a genuinely quartic
interaction (it is  just a simple harmonic
oscillator) the existence of zero-modes of (6) is somewhat
non-trivial, 
and its zero-energy eigenfunction(s) show certain features that one
may expect from a solution of the more complicated problem (involving
$M$ \underbar{and} $D$). 

To determine the spectrum of (6), diagonalize
\begin{equation}
H_D'\; :=\; 2\, f_{AA'E}\; x_{jE} \left( \delta_{\alpha 8} \;
  \delta_{\alpha'j} - \delta_{\alpha' 8}\;\delta_{\alpha j}\right)
\lambda_{\alpha A} \; \partial_{\lambda_{\alpha'A'}}
\end{equation}
by noting that if
\begin{equation}
W_{aa'}\;e_{a'}^{(\lambda)}\;=\;\lambda\;e_a^{(\lambda)} \ ,
\end{equation}
\begin{equation}
\Psi'\;=\;e_{a_1}^{(\lambda_1)} \ldots e_{a_l}^{(\lambda_l)} \;
\lambda_{a_1} \ldots \lambda_{a_l}
\end{equation}
will satisfy
\begin{equation}
W_{aa'}\; \lambda_a\;\partial_{\lambda_{a'}}\;\Psi'\;=\; \left(
  \lambda_1 +\ldots + \lambda_l\right)\;\Psi' \ .
\end{equation}
For the case at hand, (8) reads 
\begin{eqnarray}
\lambda\; e_{8A}^{(\lambda)} &=&
2\, f_{AA'E}\;x_{jE}\;e_{jA'}^{(\lambda)}\nonumber\\
\lambda\; e_{jA}^{(\lambda)} &=&
-\,2 \,f_{AA'E}\;x_{jE}\;e_{8A'}^{(\lambda)} \ ,
\end{eqnarray}
implying
\begin{equation}
\lambda^2\;e_{8A}^{(\lambda)}\;=\;+4 \,S_{AA'}\;e_{8A'}^{(\lambda)} \
,
\end{equation}
with
\begin{eqnarray}
S_{AA'} &:=& f_{ABC}\;x_{jB}\;f_{A'B'C}\; x_{jB'} \nonumber\\
&=& - \sum_{j=1}^7 \left( Ad\;X_j\right)^2_{AA'}
\end{eqnarray}
being the real symmetric (positive semi-definite, as $ \sum_j
Tr \left(\left[X_j, Z\right]^{\dagger} \ \left[ X_j, Z\right]\right) \geq 0$ )
matrix whose eigenvalues $s^{(A)} (x)\,=:\, (\omega^{(A)})^2$,
$A=1\ldots (N^2-1)$ , determine the frequencies of the harmonic
oscillators in (6).

Clearly, the ground state energy of $H_D^{(0)} \,:=\,-\,4
\partial_A\,\overline{\partial}_A + \overline{z}_A S_{AA'} z_{A'}$, is 
\begin{equation}
E_D^{(0)}\;=\;2 \left( \omega^{(1)} +\ldots+\omega^{(N^2-1)}\right) \
.
\end{equation}
Due to (11)/(12), $(W_{aa'})$ has eigenvalues $+ 2\omega_1, \ldots, +
2\omega_{N^2-1}, - 2\omega_1, \ldots, - 2\omega_{N^2-1}$ (plus $6 \cdot
(N^2-1)$ zeroes). Choosing $l \geq N^2-1$ (and $\leq 7 \cdot (N^2-1)$),
as well as $\lambda_A = - 2\omega^{(A)}$ for $A=1 \ldots N^2-1$, one
sees that the ground state energy of $H_D^{(0)}$ can be compensated by
the lowest eigenvalue of $H_D'$ -- proving that $H_D\,\Psi =0$ if 
\begin{eqnarray}
\Psi\;=\;f (x) 
&\cdot& e^{-\frac 1 2 \ \omega^{(A)}\, (x) \left( \tilde{x}_8^{\,(A)^2} +
    \tilde{x}_9^{\,(A)^2}\right)} \ \cdot \
\displaystyle\mathop{\Pi}_{A=1}^{N^2-1} 
\omega^{(A)} \nonumber\\
&\cdot& e_{a_1}^{(-2\omega^{(1)})} \;(x)\ldots
e_{a_l}^{(-2\omega^{(l)})}\;(x) \ \lambda_{a_1} \ldots \lambda_{a_l} \
,
\end{eqnarray}
with $\omega^{(l>N^2-1)}\,:=\,0$ , $\tilde{x}_8^{\,(A)} + i\, \tilde{x}_9^{\,(A)}\, :=\, R_{AA'} (x)
z_{A'}$  $(S=R^{-1} \, s\, R)$ -- and $f(x)$ subject to $\langle
\Psi, \Psi\rangle < \infty$ \ (to get a SU(N) invariant zero energy
state, one may average (15) with respect to the SU(N) action on
it). One of the interesting features of (15) is its non-trivial
dependence on $x$, in particular when two eigenvalues cross
-- or any of the $\omega^{(A)} (x)$ reach(es) zero.
It seems that the (almost trivial) supersymmetric harmonic oscillator
(6) carries quite a lot of information. 

Finally, I would like to mention the following two properties of (supersymmetric)
Hamiltonians with supercharges of the type (1), (2):

\medskip

\noindent {\bf I)} \ Let $N$ be odd, $\hat{x}_\mu\,:=\,\delta_{\mu 8}
\, x_\mu - (1-\delta_{\mu 8}) x_\mu, \mu = 1\ldots 9,$ and 
\begin{equation}
* \;:=\; \sum_m (-)^m \ \frac{\epsilon_{a_1 \ldots
    a_{\Lambda}}}{m!\,(\Lambda -m)!} \ 
\lambda_{a_1}\ldots \lambda_{a_m}\; \partial_{\lambda_{a_{\Lambda}}}
\ldots \partial_{\lambda_{a_{m+1}}} \ ,
\end{equation}
i.e.,
\begin{eqnarray}
*\;\lambda_a\;*&=&\partial_{\lambda_a} \\\nonumber
*\;\partial_{\lambda_a}\;*&=& - \lambda_a \ .
\end{eqnarray}
Then
\begin{equation}
*\;Q(\hat{x})\;*\;=\;Q^\dagger\, (x) \ ,
\end{equation}
implying 
\begin{equation}
*\;H(\hat{x})\;*\;=\;H \ .
\end{equation}
If, therefore, a zero-energy state of $H$ exists, it may be assumed to
be self-dual,
\begin{equation}
*\;\Psi (\hat{x})\;=\;\Psi (x) \ ,
\end{equation}
which means that in looking for such a state it is sufficient to only
solve $Q\,\Psi=0$ (as $Q^\dagger \Psi =0$ then follows, via (20),
(18)).

\medskip

\noindent {\bf II)} \ Let $H=-\vec{\nabla}^{\, 2} + V (\vec{x}) + H_F
(\vec{x})$ , with $V(\lambda \vec{x}) = \lambda^{2n} V(\vec{x})$ ,
$H_F (\lambda \vec{x}) = \lambda^{n-1} H_F (\vec{x})$ . $H\Psi =0$
then implies
\begin{equation}
\langle - \triangle \rangle_\Psi \;=\; \langle V\rangle_\Psi \ ,
\end{equation}
as
\begin{eqnarray}
0 &=& \langle \Psi, \;\left[ \vec{x} \cdot \vec{\nabla}, - \triangle + V +
  H_F\right]\, \Psi\rangle \nonumber \\
  &=& \langle \Psi, \;\left( 2\triangle + 2 nV + (n-1)\; H_F\right)\,
  \Psi \rangle \\
  &=& (n+1)\;\langle \Psi,\;\left( \triangle + V\right)\;\Psi\rangle \
  . \nonumber
\end{eqnarray}
In the case of (1), more detailed, local, relations can be
obtained. 

\vspace{2cm}

\noindent {\bf Acknowledgement}

\noindent I would like to thank M. Bordemann, F. Finster,
J. Fr\"ohlich, G.M. Graf, L. Motl, R. Suter and S.-T. Yau for valuable
discussions.

\vspace{3cm}

\noindent {\bf References}:
\begin{itemize}
\item[ [1$\!\!\!$] ] \ J. Hoppe; hep-th/9709217.

\end{itemize}}

\end{document}